\documentclass[%
 reprint,
superscriptaddress,
nofootinbib,
 amsmath,amssymb,
 aps,
pra,
floatfix,
]{revtex4-2}

\usepackage{graphicx}% Include figure files
\usepackage{dcolumn}% Align table columns on decimal point
\usepackage{bm}% bold math
\usepackage{dsfont}

\usepackage{xcolor,graphicx}
\usepackage{xspace}
\usepackage{amsmath}
\usepackage{braket}

\newcommand{\W}[1]{\ket{W_{#1}}\xspace}
\newcommand{\up}{\ket{\uparrow}\xspace}
\newcommand{\down}{\ket{\downarrow}\xspace}
\newcommand{\da}{\downarrow\xspace}
\newcommand{\ua}{\uparrow\xspace}
\newcommand{\Be}{\textsuperscript{9}Be\textsuperscript{+}\xspace}
\newcommand{\Mg}{\textsuperscript{25}Mg\textsuperscript{+}\xspace}
\newcommand{\Slevel}{\textsuperscript{2}S\textsubscript{1/2}\xspace}

\begin{document}

\title{Dissipative preparation of W states in trapped ion systems}

\author{Daniel C. Cole}
\email{daniel.cole@nist.gov}
\affiliation{National Institute of Standards and Technology, 325 Broadway, Boulder, CO 80305, USA}

\author{Jenny J. Wu}
\affiliation{National Institute of Standards and Technology, 325 Broadway, Boulder, CO 80305, USA}
\affiliation{Department of Physics, University of Colorado, Boulder, CO 80309, USA}

\author{Stephen D. Erickson}
\affiliation{National Institute of Standards and Technology, 325 Broadway, Boulder, CO 80305, USA}
\affiliation{Department of Physics, University of Colorado, Boulder, CO 80309, USA}

\author{Pan-Yu Hou}
\affiliation{National Institute of Standards and Technology, 325 Broadway, Boulder, CO 80305, USA}
\affiliation{Department of Physics, University of Colorado, Boulder, CO 80309, USA}

\author{Andrew C. Wilson}
\affiliation{National Institute of Standards and Technology, 325 Broadway, Boulder, CO 80305, USA}

\author{Dietrich Leibfried}
\affiliation{National Institute of Standards and Technology, 325 Broadway, Boulder, CO 80305, USA}

\author{Florentin Reiter}
\affiliation{Institute for Quantum Electronics, ETH Z\"{u}rich, Otto-Stern-Weg 1, 8093 Z\"{u}rich, Switzerland}

\begin{abstract}
We present protocols for dissipative entanglement of three trapped-ion qubits and discuss a scheme that uses sympathetic cooling as the dissipation mechanism. This scheme relies on tailored destructive interference to generate any one of six entangled W states in a three-ion qubit space. Using a beryllium-magnesium ion crystal as an example system, we theoretically investigate the protocol's performance and the effects of likely error sources, including thermal secular motion of the ion crystal, calibration imperfections, and spontaneous photon scattering. We estimate that a fidelity of $\sim$ 98 \% may be achieved in typical trapped ion experiments with $\sim$ 1 ms interaction time. These protocols avoid timescale hierarchies for faster preparation of entangled states.
\end{abstract}

\maketitle

\section{Introduction}

Dissipation arises in quantum systems through interaction with the environment and presents a challenge for applications in quantum simulation, computation, communication, and metrology. However, controlled dissipation can also be introduced and leveraged to manipulate quantum systems. Familiar examples in atomic physics include optical pumping and laser cooling. These techniques allow removal of entropy and approximate preparation of a desired pure state from an uncontrolled and unknown initial state. This cannot be accomplished by unitary operations. Recently, attention has focused on using dissipation for quantum information processing \cite{Poyatos1996,Kraus2008, Verstraete2009}, and in particular for generating entanglement. While not inherently superior to unitary entanglement-generation strategies, dissipative schemes are less sensitive to certain error mechanisms. Moreover, they allow resource states to be created and stabilized in the presence of noise so that they are available on demand. Dissipative protocols for generation and stabilization of entangled and other nonclassical states have been demonstrated in a number of systems, including macroscopic atomic ensembles \cite{Krauter2011}, trapped ions \cite{Barreiro2011, Lin2013, Kienzler2015}, and superconducting qubits \cite{Shankar2013, Kimchi-Schwartz2016, Liu2016}. Numerous proposals describe additional schemes to generate entanglement \cite{Plenio1999,Kastoryano2011,Carr2013, Rao2013,Morigi2015,Reiter2016, Shao2017}, perform error correction \cite{Pastawski2011,Reiter2017}, and initialize quantum simulators \cite{Raghunandan2020}. Broadly, the full scope over which engineered dissipation may be applied for quantum information processing is not yet clear, and practical protocols that can accomplish new tasks expand the frontier.

A first set of experiments has demonstrated preparation of steady-state entanglement using always-on couplings \cite{Lin2013, Shankar2013}. An important ingredient in these schemes is timescale hierarchies, for example the application of a strong dressing drive at rate $G$ alongside other much weaker interactions with characteristic rates $g_i\ll G$. These hierarchies protect the target state. However, the steady-state entanglement fidelity of this kind of scheme only asymptotically approaches unity as the relative strength $r=G/\max\{g_i\}$ of the dressing drive increases.  Even more importantly, timescale hierarchies limit the speed of entanglement generation, because the other interactions $g_i$ that populate the target state must be driven slowly compared to experimentally achievable rates for $G$. This puts practical limits on the speed of state preparation and the achievable fidelities in the presence of various error sources, and these limits are worse than the fundamental limit determined by $r$. Recent proposals for generation of bipartite entanglement of trapped ions \cite{Horn2018} and superconducting qubits \cite{Doucet2020} avoid imposing timescale hierarchies. As a result, they are predicted to generate entanglement more efficiently with the same experimental resources and achieve unity steady-state fidelities for finite rates in the absence of errors. In this work, we describe how couplings available in trapped-ion systems can extend these concepts for efficient dissipation engineering to the preparation of tripartite entanglement.

We present concrete schemes for the dissipative production of three-qubit W states. For single-qubit basis states $\up$ and $\down$, W states are $N$-qubit entangled states of the general form $\left(e^{i\phi_1}\ket{\uparrow\downarrow\downarrow...}+e^{i\phi_2}\ket{\downarrow\uparrow\downarrow...}+...+e^{i\phi_N}\ket{...\downarrow\downarrow\uparrow}\right)/\sqrt{N}$, where each phase $\phi_i$ is defined modulo $2 \pi$ and, as usual, the state is defined up to a global phase. W states are notable in part because their entanglement is particularly robust to particle loss\footnote{A straightforward example is the effect of measurement of a single qubit on a W state, in which case an $(N-1)$-qubit W state is preserved with probability $(N-1)/N$. This compares favorably with the same measurement on a generalized Greenberger-Horne-Zeilinger state, when entanglement is always destroyed.} \cite{Dur2000}. Unitary trapped-ion W-state generation was first demonstrated in 2005 using \textsuperscript{40}Ca\textsuperscript{+} ions  \cite{Haffner2005}, and individual addressing of the ions was crucial to that demonstration. \textit{Dissipative} preparation of W states was considered in Ref.~\cite{Ticozzi2014}, which considers quasi-local couplings that drive a multipartite quantum system into a steady state. A subsequent experiment using Hilbert-space engineering relied on global couplings and hence did not require individual qubit addressing \cite{Lin2016a}; however, it imposed a hierarchy of timescales akin to those in early dissipative state-preparation schemes. The protocols we present here require neither timescale heriarchies nor individual addressing. Instead, they exploit symmetries to enable the efficient generation of any one of six target W states in the three-qubit Hilbert space (Sec.~\ref{sec:W_state_notation}, Eqs. \ref{eq:W10}-\ref{eq:W2-}).

We structure our presentation as follows: In Sec.~\ref{sec:W_state_notation} we present a useful notation for the W states of a three-qubit system. This notation naturally describes the couplings that our schemes employ. In Sec.~\ref{sec:noiseless_subsystem} we review one possible application for these W states, the noiseless-subsystem encoding for a logical qubit. In Sec.~\ref{sec:experimental_ingredients} we describe the experimental platform in which we envision implementing these schemes. We discuss a paradigmatic system, a mixed-species, five-ion linear crystal composed of three qubit ions and two auxiliary ions of a different species for sympathetic cooling, and review the relevant features of the interactions our schemes use: resolved-sideband couplings, M\o lmer-S\o rensen spin-spin interactions, and sympathetic cooling. Section \ref{sec:schemes} describes dissipative protocols for W-state generation. In Sec.~\ref{sec:singlet_scheme_extension} we consider a natural attempt to extend the dissipative singlet-generation scheme presented by Horn \textit{et al.} \cite{Horn2018}, and discuss the failure of this scheme to generalize efficiently to more qubits. Section \ref{sec:cooling_scheme} presents a scheme for dissipative generation of W states of three ions that is more promising for experimental implementation, in which dissipation is incorporated through sympathetic cooling of the collective motion. We discuss expected error mechanisms, which indicate from simulations that dissipative W-state generation in a typical trapped-ion experiment can be achieved with $\sim$~98 \% fidelity and $\sim$~1~ms interaction time. Section \ref{sec:discussion} presents a concluding discussion.

\subsection{W state notation\label{sec:W_state_notation}}
To span the eight-dimensional Hilbert space for three qubits, we choose orthonormal basis states $\W{n_\uparrow s}$ parameterized by $n_\uparrow$, the number of qubits that are in the $\up$ state, and the chirality eigenvalue $s=0$, $+1$, or $-1$ under a chirality operator \cite{Viola2001a}:
\begin{equation}
    \chi=\frac{1}{2\sqrt{3}}\sum_{\alpha\beta\gamma}\epsilon_{\alpha\beta\gamma}\sigma_\alpha^{(1)}\sigma_\beta^{(2)}\sigma_\gamma^{(3)}. \label{eq:logZ}
\end{equation}
Here the indices $\alpha$, $\beta$, and $\gamma$ run over $x$, $y$, and $z$, $\sigma_\alpha^{(j)}$ is a Pauli matrix acting on qubit $j$, and $\epsilon_{\alpha\beta\gamma}$ is the Levi-Civita symbol.
Explicitly, the basis states $\W{n_\uparrow s}$ are:

\begin{align}
\W{00} &= \ket{\da\da\da},\label{eq:W00}\\ 
\sqrt{3}\W{10} &= \ket{\ua\da\da} + \ket{\da\ua\da} +\ket{\da\da\ua},\label{eq:W10}\\
\sqrt{3}\W{1+} &= w^*\ket{\ua\da\da} + \ket{\da\ua\da} +w\ket{\da\da\ua},\label{eq:W1+}\\
\sqrt{3}\W{1-} &= w\ket{\ua\da\da} + \ket{\da\ua\da} +w^*\ket{\da\da\ua},\label{eq:W1-}\\
\sqrt{3}\W{20} &= \ket{\da\ua\ua} + \ket{\ua\da\ua} +\ket{\ua\ua\da},\label{eq:W20}\\
\sqrt{3}\W{2+} &= w^*\ket{\da\ua\ua} + \ket{\ua\da\ua} +w\ket{\ua\ua\da},\label{eq:W2+}\\
\sqrt{3}\W{2-} &= w\ket{\da\ua\ua} + \ket{\ua\da\ua} +w^*\ket{\ua\ua\da},\label{eq:W2-}\\
\W{30} &= \ket{\ua\ua\ua},\label{eq:W30}
\end{align}
where $w = \exp{(2\pi i /3)}$ and phases are defined modulo $2\pi$. The six W states are the states with $n_\uparrow=1$ or $2$ (Eqs. \ref{eq:W10}-\ref{eq:W2-}), and we refer to the four of these states $\W{n_\uparrow\pm}$ with chirality $s=\pm1$ as chiral W states, in contrast with the achiral states for which $s=0$.

\begin{figure*}[t]
    \begin{center}
	\includegraphics[]{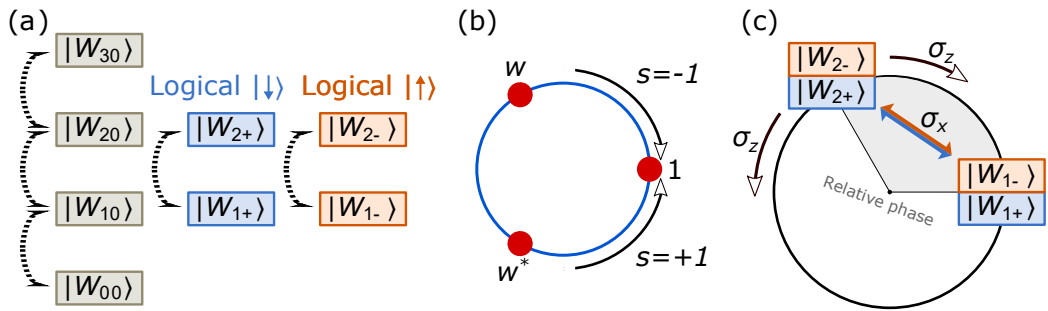}
	\caption{W states and the noiseless subsystem encoding. (a)~Couplings in the W-state basis under a global $\sigma_x$ interaction, indicated by the dotted black lines. The total-spin-$\frac{3}{2}$ ($\W{n_\uparrow0})$ and $\frac{1}{2}$ ($\W{n_\uparrow\pm}$) subspaces are closed under this interaction. The logical qubit is made up of $\down_L$, corresponding to the $s=+1$-chirality states (blue), and $\up_L$, corresponding to the $s=-1$-chirality states  (orange). (b) A graphical depiction of the chirality eigenvalues $s$, indicating the direction in which the phases of the terms $\ket{\uparrow\downarrow\downarrow}$, $\ket{\downarrow\uparrow\downarrow}$, and $\ket{\downarrow\downarrow\uparrow}$ (for $n_\uparrow=1$) traverse the unit circle. (c)~Illustration of the noiseless subsystem encoding and the action of global noise operators. Adjacent state pairs constitute decoherence-free qubits. Global $\sigma_z$ noise leads to a phase change between these decoherence-free qubits due to their different energies, displacing the $n_\uparrow=2$ qubit around the unit circle in the frame of the $n_\uparrow=1$ qubit (relative phase indicated by gray shading). Global $\sigma_x$ noise couples between the two qubits. In both cases, and for global $\sigma_y=-i\sigma_z\sigma_x$ noise, arbitrary superpositions of the logical states are preserved. \label{fig:noiseless_subsystem}}
	\end{center}
\end{figure*}

\subsection{Noiseless-subsystem qubit encoding\label{sec:noiseless_subsystem}}
Here we review one possible application for three-ion W states. In order to implement a practical quantum computer or memory, it is necessary to employ either active error correction or encodings for logical qubits that are inherently robust to noise, for which several strategies have been proposed and realized \cite{Zanardi1997,Lidar1998,Duan1998,Kielpinski2001,Langer2005,Andrews2019,Gyenis2021}.
Chiral W states $\W{n_\uparrow\pm}$ can be used in a noiseless-subsystem (NSS) qubit encoding \cite{Knill2000, Viola2001a}. This encoding generalizes a two-qubit decoherence-free subspace (DFS) encoding, which is robust against global $\sigma_z$ dephasing, to a three-qubit encoding that is insensitive to any global Pauli operator. Explicitly, the noise operator against which the encoding is protected is:
\begin{align}
    \left(\vphantom{\frac{1}{2}}\vec{\sigma}_1\cdot \vec{n}(t)\right)\otimes\mathds{1}_2\otimes\mathds{1}_3 &+ \mathds{1}_1\otimes\left(\vphantom{\frac{1}{2}}\vec{\sigma}_2\cdot \vec{n}(t)\right)\otimes\mathds{1}_3 \nonumber \\&+ \mathds{1}_1\otimes\mathds{1}_2\otimes\left(\vphantom{\frac{1}{2}}\vec{\sigma}_3\cdot \vec{n}(t)\right), \label{eq:noise_op}
\end{align}
%\begin{equation}
%    \left(\vphantom{\frac{1}{2}}\vec{\sigma}_1\cdot \vec{n}(t)\right)\otimes\mathds{1}_2\otimes\mathds{1}_3 + \mathds{1}_1\otimes\left(\vphantom{\frac{1}{2}}\vec{\sigma}_2\cdot \vec{n}(t)\right)\otimes\mathds{1}_3+ \mathds{1}_1\otimes\mathds{1}_2\otimes\left(\vphantom{\frac{1}{2}}\vec{\sigma}_3\cdot \vec{n}(t)\right), \label{eq:noise_op}
%\end{equation}
where $\vec{\sigma}_j$ is the vector of Pauli matrices and $\mathds{1}_j$ is the identity operator, each acting on qubit $j$, and the slowly varying $\vec{n}(t)$ describes the magnitude and direction of the global noise as a function of time. Examples of NSS encodings include an exchange-only quantum dot spin qubit~\cite{DiVincenzo2000, Russ2017, Andrews2019} and the encoding we discuss below, which has been realized and explored in nuclear magnetic-resonance systems \cite{Viola2001z, Fortunato2003}. 

In a subsystem encoding, a qubit is encoded in a tensor factor of a Hilbert \mbox{(sub-)space}. A noiseless subsystem encoding is realized when the qubit degree of freedom does not couple to the global environment; only the remaining ``gauge" degrees of freedom do. In the NSS encoding that we consider, a qubit is encoded in the total-spin-$\frac{1}{2}$ subspace that is spanned by the chiral W states. States within this subspace can be factored into joint eigenstates of a gauge $S_z$ degree of freedom with eigenvalues $\pm\frac{\hbar}{2}$ (related to the label $n_\uparrow$) and a chiral qubit degree of freedom with eigenvalues $\pm1$. Global noise of the form described by Eq. \ref{eq:noise_op} couples only to the former, so that the qubit is unaffected if the two degrees of freedom are separable.

Concretely, we associate the logical state $\down_L$ with chirality $s=+1$ and $\up_L$ with chirality $s=-1$. With this choice, the pair $\W{1+}$, $\W{1-}$ and the pair $\W{2+}$, $\W{2-}$ each constitute a DFS qubit that is protected from global $\sigma_z$ noise, and global $\sigma_x$ and $\sigma_y$ noise map between these qubits. Arbitrary logical states are protected against arbitrary global noise as defined in Eq. \ref{eq:noise_op} when they are encoded as:
\begin{align}
    \alpha_L\down_L+\beta_L\up_L &\longrightarrow G_1\left(\alpha_L\W{1+}+\beta_L\W{1-}\right) \nonumber \\&+ G_2\left(\alpha_L\W{2+}+\beta_L\W{2-}\right),
\end{align} 
%\begin{equation}
%    \alpha_L\down_L+\beta_L\up_L \longrightarrow G_1\left(\alpha_L\W{1+}+\beta_L\W{1-}\right) + G_2\left(\alpha_L\W{2+}+\beta_L\W{2-}\right),
%\end{equation} 
which can be factored into the gauge and qubit degrees of freedom as:
\begin{align}
    \alpha_L\down_L+&\beta_L\up_L \longrightarrow \left(G_1\ket{n_\uparrow=1}+G_2\ket{n_\uparrow=2}\right)\nonumber \\ &\otimes\left(\alpha_L\ket{s=+1}+\beta_L\ket{s=-1}\right). \label{eq:tensorfactor}
\end{align} 
%\begin{equation}
%    \alpha_L\down_L+\beta_L\up_L \longrightarrow \left(G_1\ket{n_\uparrow=1}+G_2\ket{n_\uparrow=2}\right)\otimes\left(\alpha_L\ket{s=+1}+\beta_L\ket{s=-1}\right). \label{eq:tensorfactor}
%\end{equation}
In the above, $|\alpha_L|^2 + |\beta_L|^2=1$ and $|G_1|^2 + |G_2|^2=1$. Global noise only changes the coefficients $G_i$. The separability represented in Eq. \ref{eq:tensorfactor} is crucial. For example, $(\W{1+}+\W{2-})/\sqrt{2}$ is not a valid encoding of $(\down_L+\up_L)/\sqrt{2}$ because the qubit and gauge degrees of freedom are entangled, and the logical state is not preserved under global $\sigma_z$ noise. The encoded logical qubit can be manipulated using suitable operators \cite{Viola2001a}, with the chirality operator $\chi$ (Eq. \ref{eq:logZ}) acting as the single-qubit logical $Z$ operator. This NSS encoding is illustrated schematically in Fig. \ref{fig:noiseless_subsystem}.

Implicit in Eq. \ref{eq:noise_op} is a choice about the orientation of each qubit's Bloch sphere. The NSS can protect against either spatially uniform noise or a noise field that has spatially uniform intensity but a phase that varies with position, such as a laser field. In this case the orientation of each qubit's Bloch sphere can be fixed by stipulating the direction of the noise field at a given time so that Eq. \ref{eq:noise_op} applies.

One application that would use the full power of the NSS's robustness against general uniform noise would be to preserve quantum information during application of global control pulses (the `noise' field) that affect all physical qubits in a quantum register, including those making up the logical NSS qubits. For example, the NSS encoding can protect against over- and under-rotation errors in global pulses used for dynamical decoupling \cite{Viola1999}, which themselves can reduce decoherence due to uniform \textit{or} spatially varying fluctuations in, e.g., a quantization field. In this way, the NSS encoding can be used as part of a strategy for mitigating nonuniform noise. The NSS encoding may also be used to realize a reference-frame-free qubit for quantum communication \cite{Bartlett2003}.

\section{Experimental ingredients\label{sec:experimental_ingredients}}

The dissipative protocols we present make use of standard techniques for ion-trap experiments: resolved sideband transitions, effective spin-spin couplings driven by M\o lmer-S\o rensen-type interactions, and sympathetic cooling with an auxiliary species. In this section we introduce an example system that could be used, a linear crystal of beryllium-9 (\Be) and magnesium-25 (\Mg) ions, and then review the relevant features of these techniques.

\subsection{Mixed-species ion crystal\label{sec:MBBBM}}
We consider a system in which three qubit ions and two auxiliary ions in the same trapping potential form a one-dimensional Coulomb crystal along the axis of weakest confinement \cite{James1998}. The Coulomb interaction leads to 15 quantized normal modes of collective motion. We assume that the ten modes of radial motion along the dimensions of tighter confinement factor from the five axial modes, and that the interactions we discuss below couple only to the axial degrees of freedom. For concreteness, we discuss an ion crystal of the form \Mg-\Be-\Be-\Be-\Mg (`MBBBM'), but these schemes could be realized with a variety of qubit and auxiliary species. Confining this crystal in a parity-symmetric axial potential provides symmetric crystal geometry and axial mode eigenvectors that are either symmetric or antisymmetric under parity \cite{Morigi2001}. The $l$\textsuperscript{th} eigenvector can be described by entries $z_j^{(l)}$ that specify the participation of ion $j$ in the mode's zero-point motion in units of length.

\subsection{Resolved sideband transitions \label{sec:sidebandtransitions}}

When ions of different species with optical transitions of different wavelengths are used for qubits and for cooling, it is typically possible to address the two species independently. In the Lamb-Dicke approximation and the interaction picture for the qubit and the vibrational levels, and neglecting rapidly oscillating terms, a blue-sideband interaction resonant with mode $l$ of the ion crystal driven by addressing only the qubit ions can be described by the Hamiltonian \cite{Wineland1998, Lee2005}:
%\begin{align}
%    H_{sb} =  a^\dagger\sum_{j=j_\mathrm{qubit}}(\pm)_j\frac{|\Omega_j|}{2} \sigma_+^{(j)} &e^{i (\Delta k \cdot Z_{0,j}+\Delta\phi)} \nonumber \\&+ H.c., \label{eq:HSB}
%\end{align}
\begin{equation}
    H_{sb} =  a^\dagger\sum_{j=j_\mathrm{qubit}}(\pm)_j\frac{|\Omega_j|}{2} \sigma_+^{(j)} e^{i (\Delta k \cdot Z_{0,j}+\Delta\phi)}+ H.c., \label{eq:HSB}
\end{equation}
and for a red-sideband interaction the creation operator $a^\dagger$ for mode $l$ is replaced by the annihilation operator $a$. Here $\sigma_+^{(j)} = \up_j\bra{\downarrow}_j$ and $|\Omega_j|$ is the magnitude of the effective Rabi frequency of the interaction for the $j$\textsuperscript{th} ion. For \Be ground-state hyperfine qubits, the Rabi frequency $\Omega_j$ is proportional to the electric field strengths of the two laser beams driving a stimulated Raman transition transition and to $\eta_j^{(l)}=\Delta k \cdot |z_j^{(l)}|$, the Lamb-Dicke parameter quantifying the $j$\textsuperscript{th} ion's participation in the interaction. Here $\Delta k$ is the axial projection of the difference wavevector between the Raman beams. In addition to a global phase coming from the phase difference $\Delta\phi$ between the laser beams, there is a $j$-dependent phase represented by $e^{i\phi_j}=(\pm)_j e^{i \Delta k \cdot Z_{0,j}}$, where $Z_{0,j}$ is the equilibrium axial coordinate of ion $j$ and $(\pm)_j=\mathrm{Sign}\left(z_j^{(l)}\right)$. We can control these phases by adjusting the ions' relative positions along the trap axis or by adjusting the magnitude of $\Delta k$. For particular values of these phases, the sideband interaction can induce definite changes $\delta s$ in the chirality eigenvalue $s$.  We associate $\delta s$ with the coupling:

\begin{equation}
\W{00}\leftrightarrow\W{1(\delta s)}\leftrightarrow \W{2 (-\delta s)} \leftrightarrow \W{30},
\end{equation}
and then, for example, $ \delta s=+1$ for $\phi_j=2\pi/3\cdot j$ for the MBBBM crystal.

Any non-uniformity of $|\Omega_j|$ across the qubit ions will introduce errors in the schemes we describe below. Important sources of non-uniformity are different participation amplitudes $|z_{j}^{(l)}|$ in the mode that is driven by the sideband interaction and laser beam intensity differences across the ions.  The latter effect can be mitigated by illuminating the ions with beams that have approximately uniform intensity over the extent of the ion crystal, and we address the issue of mode participation in Sec.~\ref{sec:anh_pot}.

\subsection{M\o lmer-S\o rensen effective spin-spin interaction \label{sec:MSgates}}

In the M\o lmer-S\o rensen (MS) interaction \cite{Sorensen1999,Molmer1999,Solano1999,Sorensen2000,Milburn2000}, an effective spin-spin coupling is driven by red and blue sidebands with opposite detunings $\pm\delta$ from resonance with mode $l$. Assuming that the Rabi frequency magnitudes $|\Omega_j|$ are uniformly equal to $\Omega$ for the qubit ions (and zero for the auxiliary ions), the near-resonant terms in the Hamiltonian are:
\begin{align}
    H_{MS} =  \frac{\Omega}{2} &\sum_{j=j_\mathrm{qubit}} (\pm)_j \sigma_+^{(j)} \left[\vphantom{\frac{1}{2}}a^\dagger\ e^{i(\Delta\phi_b+ \Delta k_b \cdot Z_{0,j}+\delta t)} \right.\nonumber \\&\left. + a e^{i(\Delta\phi_r + \Delta k_r \cdot Z_{0,j}-\delta t)} \vphantom{\frac{1}{2}}\right] + H.c., \label{eq:HMS}
\end{align}
%\begin{equation}
%    H_{MS} =  \frac{\Omega}{2} \sum_{j=j_\mathrm{qubit}} (\pm)_j \sigma_+^{(j)} \left[\vphantom{\frac{1}{2}}a^\dagger\ e^{i(\Delta\phi_b+ \Delta k_b \cdot Z_{0,j}+\delta t)} + a e^{i(\Delta\phi_r + \Delta k_r \cdot Z_{0,j}-\delta t)} \vphantom{\frac{1}{2}}\right] + H.c., \label{eq:HMS}
%\end{equation}
Here $\Delta k_b$, $\Delta k_r$, $\Delta\phi_b$, and $\Delta\phi_r$ represent the corresponding quantities from Eq. \ref{eq:HSB} that describe the pair of beams that drives each sideband. For interaction times $T = 2\pi m/\delta$ for integer $m$, the MS interaction realizes the propagator for a spin-spin Hamiltonian on the qubit ions, returning the motion to its initial state. 

\begin{figure}
    \begin{center}
	\includegraphics[]{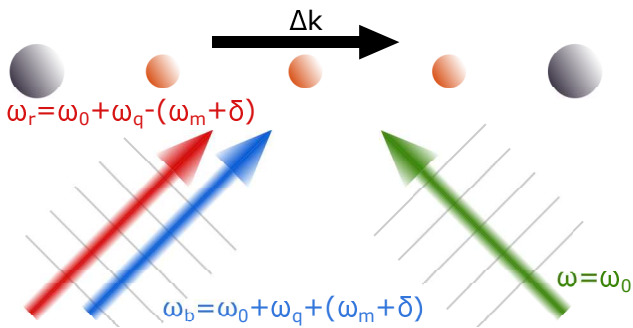}
	\caption{An illustration of the phase-sensitive geometry for the M\o lmer-S\o rensen interaction, in which the spacing between the ions or the magnitude of $\Delta k$ may be used to control the induced chirality change. The \Be qubit ions are indicated by small orange spheres, and the \Mg auxiliary ions are indicated by large gray spheres. The $k$ vectors for laser beams of three frequencies are depicted, with the red and blue beams (left) driving red and blue sideband transitions when combined with the green beam (right).  Each beam has a waist much larger than the ion crystal, so that the qubit ions are illuminated approximately uniformly. Equiphase lines are indicated perpendicular to the beam $k$ vectors. The magnitudes of the $k$ vectors are nearly the same for all three beams because $\omega_0\gg\omega_q\gg\omega_m+\delta$. The $\Delta k$ vector is approximately axial and is nearly the same for the red/green and blue/green beam pairs. \label{fig:geometry}}
	\end{center}
\end{figure}

The exact propagator that is obtained depends both on the ratio $\Omega/\delta$ and on the experimental geometry. We consider an implementation of the ``phase sensitive" geometry for driving stimulated Raman transitions (see Ref. \cite{Lee2005}), which we illustrate in Fig. \ref{fig:geometry}. In this configuration, copropagating beams at frequencies $\omega_o + \omega_q \pm (\omega_m + \delta$) intersect on the ions with an orthogonal beam of frequency $\omega_o$, such that $\Delta\vec{k}$ is axial for each orthogonal pair. Here $\omega_o$ is an optical frequency detuned from a transition between the ground-state qubit manifold and an electronic excited state, $\omega_q$ is the qubit frequency, and $\omega_m$ is the frequency of the $l$\textsuperscript{th} motional mode, upon which $a$ and $a^\dagger$ act. The phases and axial wavenumbers for the beam pairs are approximately the same: $\Delta\phi_r \simeq \Delta\phi_b=\Delta\phi$ and $\Delta k_r \simeq \Delta k_b$.

In this configuration, population transfer is maximized for $\delta = 2 \Omega$ and interaction time $T=2\pi/\delta$. For red- and blue-sidebands driving chirality change $-\delta s$, the MS interaction couples $\W{00}\leftrightarrow\W{2 (\delta s)}$. In matrix notation, the action of the gate on the state vector $\left(\W{00},\W{1(-\delta s)},\W{2\delta s}, \W{30}\right)^T$ is:

\begin{gather}
 U
 =e^{i\pi/8}
  \begin{pmatrix}
   \frac{1}{2} & 0 & -\frac{\sqrt{3}}{2}u^* & 0  \\
   0 & -\frac{1}{2} & 0 & -\frac{\sqrt{3}}{2}u^*   \\
   -\frac{\sqrt{3}}{2}u & 0 & -\frac{1}{2} & 0   \\
   0 & -\frac{\sqrt{3}}{2}u & 0 &  \frac{1}{2}   \\
   \end{pmatrix},
\end{gather}
where $u=e^{2i \Delta \phi}$. The subspace spanned by the four other basis states is left invariant. For $s=0$, the chiral W states are left invariant. For $s=\pm1$, two chiral and two achiral W states are left invariant. A maximum of 75~\% population transfer can be achieved between the pairs of the remaining states separated by $\Delta n_\uparrow = \pm2$.

\subsection{Achieving uniform $|\Omega_j|$ using anharmonic potentials\label{sec:anh_pot}}

\begin{figure}
    \begin{center}
	\includegraphics[width=\linewidth]{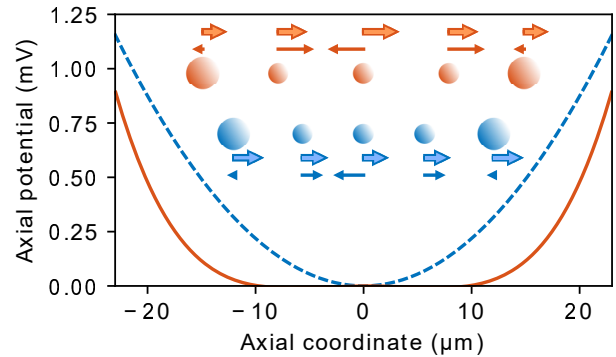}
	\caption{Two possibilities for the static axial trapping potential for the MBBBM crystal described in the text. A harmonic potential is shown in dashed blue, and a potential with significant but practical anharmonicity is shown in solid orange. The confinement is chosen so that the frequency of in-phase oscillation is 0.81 MHz for both crystals. The ion crystals at equilibrium for the anharmonic (top, orange) and harmonic (bottom, blue) potentials are shown, with the beryllium qubit ions corresponding to the smaller spheres, along with relative participation vectors for the lowest-frequency (in-phase) mode of motion (thicker, light-filled arrows) and the highest-frequency mode of motion (dark arrows) for each crystal. The zero-point amplitudes of the in-phase mode in the harmonic potential are $z^{(1)}_\mathrm{harm}=(9.56, 7.89, 7.62, 7.89, 9.56)$ nm, and for the highest-frequency mode of the anharmonic potential, with 2.56 MHz frequency, they are $z^{(5)}_\mathrm{anharm}=(-1.38, 8.35, -8.35, 8.35, -1.38)$ nm. Because the latter has reduced center-of-mass motion, it is expected to have a heating rate due to excitation by global fields that is 5.4 \% that of hypothetical in-phase motion at the same frequency (see \ref{app:crystal_modes}). \label{fig:MBBBMcrystal}}
	\end{center}
\end{figure}

As discussed in Sec.~\ref{sec:sidebandtransitions} and assumed in Sec.~\ref{sec:MSgates}, it is important that $|\Omega_j|$ is uniform across the qubit ions. The lowest-frequency, in-phase axial mode for the MBBBM crystal has nearly uniform participation. However, this mode is especially susceptible to heating by homogeneous electric field noise \cite{King1998} because of the in-phase motion and the inverse scaling of the heating rate with frequency \cite{Brownnutt2015}. Therefore, it may be better to use one of the higher-frequency modes in which some of the ions oscillate out of phase. Uniform participation of the qubit ions in one of these modes can be achieved by introducing some controlled anharmonicity to the trapping potential, as we illustrate in Fig. \ref{fig:MBBBMcrystal}. The anharmonicity assumed here can be realized in existing ion traps, e.g., the one described in Ref. \cite{Blakestad2010} with $\sim$~150~$\mu$m ion-electrode distance. We discuss this issue in greater depth for a variety of ion species and configurations in \ref{app:crystal_modes}.

\subsection{Sympathetic cooling}
In Sec.~\ref{sec:cooling_scheme} we use sympathetic cooling \cite{Larson1986, Kielpinski2000} as a dissipation mechanism that allows one-way population flow into a target W state. Resolved-sideband cooling on auxiliary ions can be used to ground-state cool any crystal mode in which they participate without disturbing the qubit states, provided that the radiation used to manipulate the auxiliary species is far-detuned from the qubit-species transitions. 

\section{Dissipative preparation of W states\label{sec:schemes}}

In this section we present protocols for dissipative preparation of W states. First, we consider attempts to extend the rapid, symmetry-based dissipative singlet-generation scheme presented by Horn \textit{et al.} in Ref. \cite{Horn2018} to more ions. This scheme incorporates dissipation through spontaneous decay from an electronic excited state. Direct extension is not promising because the rate at which the target state can be populated decreases exponentially with the number of ions. We then present a more promising approach in which dissipation is provided by sympathetic cooling, and we provide an analysis of likely error sources.

The simulation results we present are obtained by numerical simulation of the quantum-mechanical master equation in Lindblad form:
\begin{equation}
\partial_t \rho = -i[H,\rho] +\mathcal{L}_D\rho,
\end{equation}
where $\rho$ is the system's density matrix, $H$ is the Hamiltonian (of the form prescribed by Eqs. \ref{eq:HSB} and \ref{eq:HMS} for a given interaction), and $\mathcal{L}_D$ is the Lindblad dissipator:
\begin{equation}
\mathcal{L}_D \rho = \sum_k \left[L_k \rho L_k^\dagger - \frac{1}{2}\left\{\rho, L_k^\dagger L_k\right\}\right],
\end{equation}
where $L_k$ are the jump operators for the system that represent deliberate controlled dissipation and also undesirable dissipation mechanisms such as heating of the crystal. Our simulations are carried out with QuTiP's master equation solver \cite{Johansson2013}.

\subsection{A natural extension of a scheme for dissipative singlet preparation\label{sec:singlet_scheme_extension}}
The scheme presented by Horn \textit{et al.} \cite{Horn2018} generates a singlet state $\ket{S} =\left(\ket{\uparrow\downarrow} - \ket{\downarrow\uparrow}\right)/\sqrt{2}$ using engineered decay through an electronic excited state. We briefly review the basic principle: a third stable ground-state level $\ket{r}$ is pumped to an electronic excited state, which then decays back to $\up$, $\down$, or $\ket{r}$, randomly populating the nine-dimensional two-qutrit Hilbert space. By implementing global carrier and motion-adding interactions that act on the qubits as $\ket{\downarrow\downarrow}\leftrightarrow\left(\ket{\downarrow\uparrow} + \ket{\uparrow\downarrow}\right)/\sqrt{2}\leftrightarrow\ket{\uparrow\uparrow}$, and a motion-subtracting coupling from $\up$ to $\ket{r}$, population can be cycled through the excited state until it becomes trapped in $\ket{S}\otimes\ket{n=0}$, which is the sole dark state of the full system dynamics.
A specific set of levels must be chosen for this scheme so that the excited state can decay only to $\up$, $\down$, or $\ket{r}$. For the concrete case of the \Be \Slevel manifold, one possible choice is to use $\ket{F = 2, m_F = 2} = \ket{\downarrow}$, $\ket{F=1, m_F = 1} = \ket{\uparrow}$, and $\ket{F=2, m_F = 1} = \ket{r}$, with a repumper laser driving the transition to the excited state $\ket{r}\rightarrow\ket{^2 P_{1/2}, F=2, m_F = 2}$.

Naively applying the same interactions on a three-ion system results in continuous population cycling that leads to a steady-state distribution of population in a mixed state. So far we have not identified a promising three-qubit protocol that uses spontaneous emission as the dissipation mechanism. However, we briefly present one example scheme; this scheme's flaw is that it is prohibitively slow.

\begin{figure*}
    \begin{centering}
	\includegraphics[]{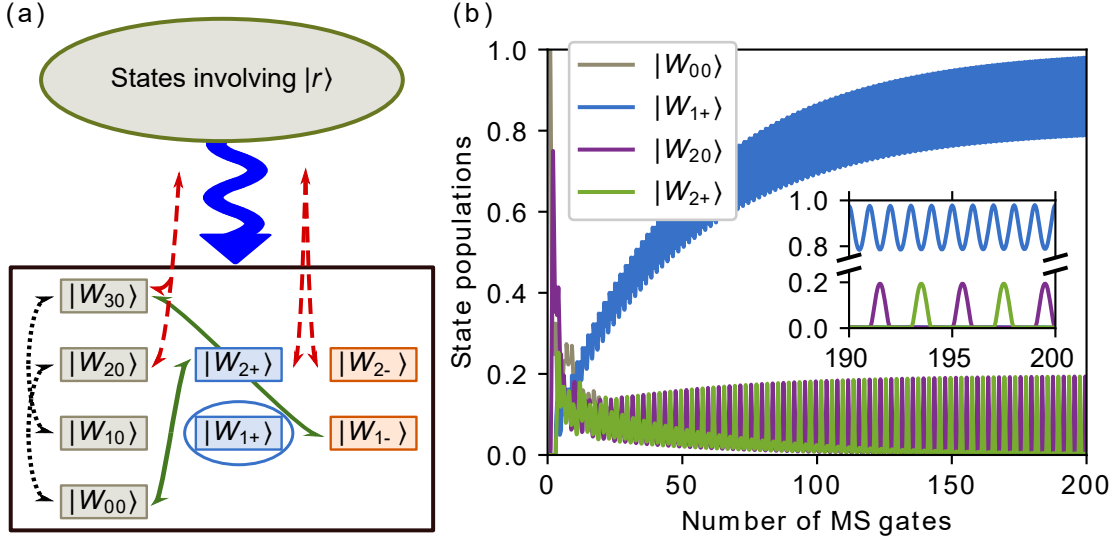}
	\caption{A scheme using spontaneous emission for generation of a W state, here $\W{1+}$. (a) Population transfer is driven by interactions including an achiral qubit MS interaction (dotted black lines), a chiral qubit MS interaction (solid green lines), an $\up$-$\ket{r}$ MS interaction (dashed red lines), and effective decay out of states involving $\ket{r}$ (snaking blue line) back to the qubit space (black box). Population becomes trapped in the target state $\W{1+}$ (blue oval). (b)~Simulated dynamics. From an initial state $\ket{\downarrow\downarrow\downarrow} = \W{00}$, M\o lmer-S\o rensen gates and repumping cycle population through an excited state, which randomly populates the qubit space upon decay. The full pumping process leaves the target state $\W{1+}$ invariant, so population accumulates there. After the 100\textsuperscript{th} repumping step (200 MS gates), a fidelity of 98.0 $\%$ is achieved. Since no error mechanisms are included in this simulation, the fidelity will continue increasing aysmptotically to 1. Inset: Zoomed view of the dynamics, showing that population is transiently removed from $\W{1+}$ during gate operation but is returned at the end of the gate. During the qubit-manifold MS gates, the states $\W{20}$ and $\W{2+}$ are briefly populated. During the $\up$-$\ket{r}$ MS gates, states involving $\ket{r}$ are transiently populated (not shown).} \label{fig:MSscheme}
	\end{centering}
\end{figure*}

We replace both the global qubit-manifold drive and the drive from $\up$ to the auxiliary state $\ket{r}$ of the singlet scheme with MS interactions. For the drive to $\ket{r}$, an MS interaction enforces pair-wise transitions and prevents one-up states from coupling to states that include $\ket{r}$. Within the qubit manifold, an MS interaction achieves for the three-qubit Hilbert space what a global interaction achieves for the two-qubit Hilbert space: it leaves the target state invariant, but makes connections between other states. Unfortunately, the MS interaction leaves too many states invariant. We can work around this problem by iterating over application of an achiral qubit MS gate, driving population from $\up$ to $\ket{r}$ with an MS gate, and repumping, and then applying a \textit{chiral} qubit MS gate and carrying out the repumping process again. The chirality change $\delta s=\pm1$ of the chiral MS gate is chosen to leave the desired chiral W state invariant. The alternating chirality of the MS gates can be achieved by periodic adjustment of the spacing of the ion crystal or the magnitude of $\Delta k$.

In Fig. \ref{fig:MSscheme}, we present an illustration of the concept and a simulation of the dynamics for a chiral qubit MS gate with $\delta s = +1$. In the simulation, we use jump operators to model the repumping as: 
\begin{equation}
L_q^{(j)} = \sqrt{b_q \Gamma}\ket{q}_j\bra{r}_j,
\end{equation}
where $q = \uparrow$, $\downarrow$, or $r$ indexes over the states, $j$ indexes over the ions, $\Gamma$ is the rate of effective decay out of $\ket{r}$, and $b_\uparrow=5/12$, $b_\downarrow=1/3$, and $b_r=1/4$ approximately describe the branching of the decay out of the excited state $\ket{e}$ \cite{Lin2013}. These jump operators are obtained via adiabatic elimination of $\ket{e}$ \cite{Reiter2012}. Since the duration of the MS gates is expected to be the speed bottleneck for this scheme, the simulations implement the repumping step for long enough that $\ket{r}$ is effectively fully depleted, and the repumping dynamics are not shown explicitly. As shown in Fig. \ref{fig:MSscheme}b, the dissipative dynamics lead to trapping of population in the state $\ket{W_{1+}}$ that is invariant under these combined dynamics.

Fig. \ref{fig:MSscheme} shows that hundreds of MS gates are needed to approach 100 \% fidelity, even in the absence of errors and imperfections. This illustrates a fundamental limitation of the extension of the singlet scheme to larger numbers of qubits: as the dimension of the qubit space increases exponentially, the rate at which the target state is populated through random decay out of the electronic excited state decreases correspondingly. Therefore, protocols that rely on randomly populating the target state scale inherently poorly.

\subsection{Dissipative preparation of chiral W states using sympathetic cooling\label{sec:cooling_scheme}}

Now we present a more promising protocol that incorporates dissipation through sympathetic cooling, which can convert unitary, periodic dynamics to one-way population flow. Unfortunately, it is not possible to generate a W state by simply evolving out of the state $\ket{\downarrow\downarrow\downarrow}$ with a coupling $\down\leftrightarrow\up$. Although a state $\ket{W_{1 s}}$ can be populated under this coupling, the overall state remains separable as population flows to states with increasing $n_\uparrow$. This flow can be truncated by preparing a Fock state $\ket{n=1}$ and then operating on a red sideband $\ket{\downarrow\downarrow\downarrow, n = 1} \leftrightarrow\ket{W_{1 s}, n = 0}$ such that a W state is created under unitary evolution \cite{Haffner2005}, but implementing this strategy is not straightforward without individual addressing.

The protocol we present gets around these problems by iterating over two steps: First, some fraction of the population less than 100 \% is transferred to the target W state through operations that also induce a motional transition $n\rightarrow n+1$ on a collective mode. Second, sympathetic cooling removes phonons from this mode. As this mode approaches its ground state, these steps can be repeated without removing population that has already reached the target state, so that population can accumulate there over several iterations. 

From an initial state $\W{00}\otimes\ket{n=0}$, the basic steps proceed as follows: An achiral MS gate transfers 75~\%  of the population to $\W{20}$. A $\delta s = \mp1$ red sideband interaction then moves this population to the state $\W{1\pm} \otimes\ket{n=1}$. When the ion crystal is sympathetically cooled, the acquired phonon is removed. In practice, the cooling and sideband interaction can be applied simultaneously. Repeated iteration of these steps pumps population to the target state $\W{1\pm}\otimes\ket{n=0}$, from which it cannot escape: The chiral W states are invariant under the achiral MS gate, the $\delta s = \mp1$ sideband does not couple $\W{00}\leftrightarrow\W{1\pm}$, and $\W{1 \pm}\otimes\ket{n=0}$ is prevented from evolving to $\W{20}$ under the $\mp1$-chirality red sideband because it is in the ground state of the motion.

\begin{figure*}
    \begin{centering}
	\includegraphics[]{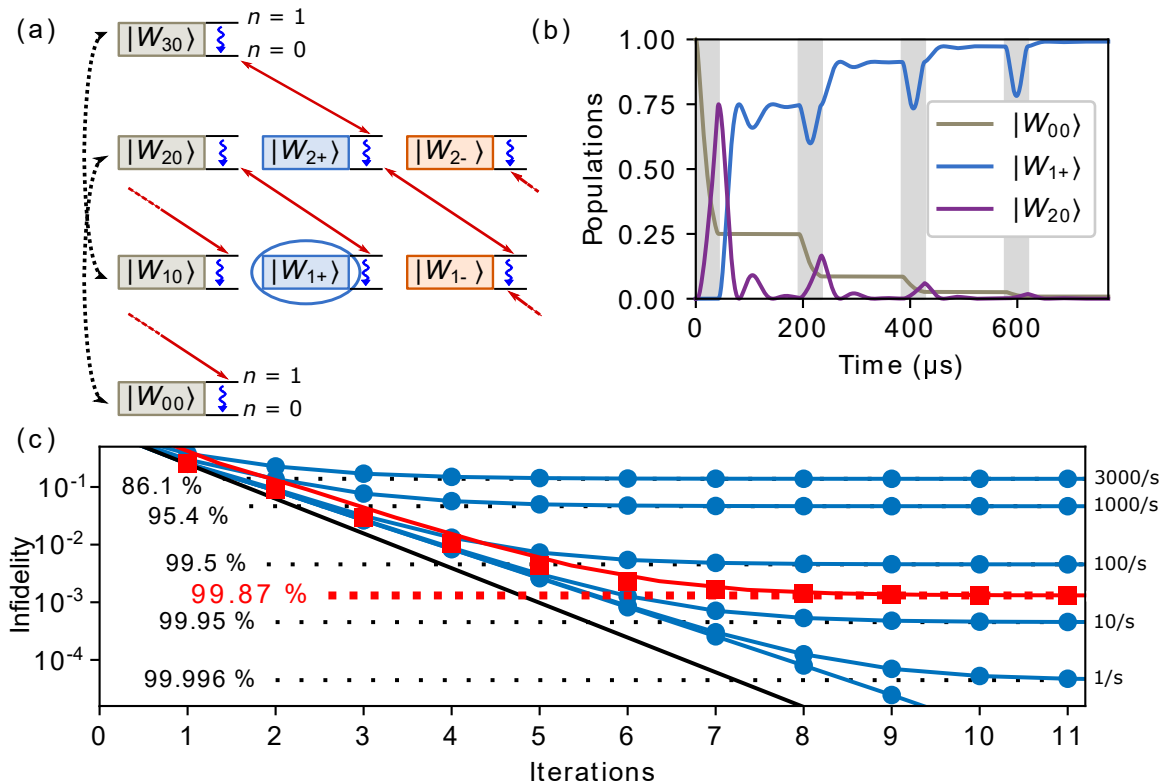}
	\caption{Illustration and numerical investigation of a protocol using sympathetic cooling to generate a W state, here $\W{1+}$. (a)~Population transfer is driven by interactions including an achiral MS interaction (dotted black lines), a chiral red sideband interaction (straight red lines; broken lines indicate couplings that wrap around the edges of the diagram), and sympathetic cooling (snaking blue lines). Only the $n=0$, $1$ normal mode Fock states are displayed next to the qubit states for simplicity. Population becomes trapped in the target state $\W{1+}$ (blue oval). (b)~Simulated dynamics, beginning from an initial qubit state $\ket{\downarrow\downarrow\downarrow} = \W{00}$ and the motional ground state. A M\o lmer-S\o rensen gate applied for $\sqrt{2}\times$30 $\mu$s (gray shading) moves population to $\W{20}$, after which simultaneous application of sympathetic cooling and a chiral sideband interaction move this population to $\W{1+}$. (c)~Target W state populations at the end of the sideband/cooling step, shown for heating rates of zero, 1 (.0052), 10 (.052), 100 (.52), 1000 (5.2), and 3000 (16) quanta/sec (quanta/iteration); higher heating rates correspond to larger infidelity. Populations obtained after twenty iterations are indicated next to the curves and by dotted lines. Also shown is the infidelity for the case of zero heating rate but with $\sim3.4$ \% sideband Rabi rate $\Omega_j$ imbalances corresponding to the lowest frequency (in-phase) mode of the MBBBM crystal in a harmonic confining potential (red squares). Solid black line indicates simple fidelity estimate of $1-1/4^n$ after $n$ iterations. If the duration of the sideband/cooling step is increased the zero heating-rate performance of the scheme follows this approximation more closely, but the total duration of the scheme increases. } \label{fig:cooling_scheme_basic}
	\end{centering}
\end{figure*}

Fig. \ref{fig:cooling_scheme_basic} presents an illustration of the concept and a numerical investigation of the dynamics for chirality change $\delta s=+1$. As can be seen from Fig. \ref{fig:cooling_scheme_basic}a, there is a path to the target state $\W{1+}$ from any of the eight basis states. In fact, the protocol will move all population in the qubit space to the target state, as we verify by simulating the dynamics beginning from each of 64 states in a complete basis for the space of three-qubit density matrices \cite{Chuang1997}. Moreover, it is straightforward to adjust the interactions to generate any of the six W states defined in Eqs. \ref{eq:W10}-\ref{eq:W2-}.

The simulation shown in Fig. \ref{fig:cooling_scheme_basic}b iterates between application of an achiral MS pulse and simultaneous application of the chiral sideband and sympathetic cooling. For concreteness we assume that a sideband transition on both the qubit and the auxiliary\footnote{For the specific case of the MBBBM crystal described in Sec. \ref{sec:MBBBM}, the significantly smaller relative participation of the magnesium ions in the driven mode can be compensated for by using Raman beams with smaller detuning for sympathetic cooling, because a higher photon scattering rate can be tolerated on the auxiliary species.} ions can be driven in 30 $\mu$s. Therefore, the time for the MS gate is $\sqrt{2}\times$30 $\mu$s in the phase-sensitive geometry, as half the optical power in one of the Raman beams contributes to driving each MS sideband and the sideband Rabi rate scales with the field amplitude \cite{Wineland1998}, and sympathetic cooling can extract phonons at a maximum rate of 2/30 $\mu$s. The jump operator that implements sympathetic cooling with rate $\kappa$ is $L = \sqrt{\kappa}a$. Interactions of alternating chirality can be implemented by periodically changing the spacing of the ion crystal or the magnitude of $\Delta k$.

If the population transfer from $\W{20}$ to $\W{1+}$ is complete, the fidelity after $n$ iterations is $1-\frac{1}{4^n}$. In practice, this depends on the details of the sideband/cooling step. Our simulations apply the sideband/cooling interaction for 150 $\mu$s, which achieves 99~\% one-way population transfer from $\W{20}\otimes \ket{n=0}$ to $\W{1+}\otimes \ket{n=0}$. 

\subsubsection{Effect of motional occupation on preparation fidelity} This scheme relies on occupation of the motional ground state to prevent population from leaking out of the target state during the simultaneous sideband/cooling step. Therefore, heating of the driven mode will reduce the fidelity, and this is expected to be the limiting error mechanism if the heating is significant. On the other hand, since sympathetic cooling is built in, the final fidelity is asymptotically insensitive to the initial motional state. In Fig. \ref{fig:cooling_scheme_basic}c we present numerical calculations of the fidelity for several heating rates. As described previously, this error mechanism can be mitigated by using a mode that has anti-parallel participation vectors for different ions, which reduces its sensitivity to global fields.

Heating of the motion with rate $\Gamma$ is implemented in the master equation with two collapse operators $\sqrt{\Gamma}a$ and $\sqrt{\Gamma}a^\dag$ \cite{Wineland1998,Sorensen2000}. When this is combined with cooling at rate $\kappa$, implemented with collapse operator $\sqrt{\kappa}a$, the steady-state mean phonon number is $\bar{n}=\Gamma/\kappa$. However, when $\Gamma$ is determined experimentally by measuring the change in phonon population after a variable probe delay \cite{Turchette2000}, this ratio may predict a limit $\bar{n}_{GSC}$ for ground-state cooling (GSC) that is far below what is experimentally achieved. This is because the sideband cooling process itself includes sources of heating, such as off-resonant excitation of motion-adding transitions and recoil heating during scattering of repump photons. To address this, in Table \ref{table:GSC} we present additional calculations of the fidelity in the presence of heating applied only during the sympathetic cooling step to phenomenologically model a given GSC limit. These results show that the scheme benefits from excellent ground-state cooling, but does not require unreasonably low $\bar{n}_{GSC}$.

\begin{table*}[t!]
\begin{centering}
	\begin{tabular}{|c|c|c|}
		\hline
		$\bar{n}_{GSC}$ & Phenomenological heating rate $\Gamma$ (s$^{-1})$ & Fidelity ($\%$)\\\hline
		0.001&  67 & 99.8 \\\hline
		0.003 & 200 & 99.4 \\\hline
		0.01 & 667 & 98.0 \\\hline
		0.03 & 2000 & 94.4 \\\hline
		0.1 &  6667 & 84.0 \\\hline
		0.3 &  20000 & 65.2 \\\hline
	\end{tabular}
	\caption{W-state fidelity after twenty iterations as a function of phenomenological heating rate $\Gamma$ applied to model a given ground-state cooling limit $\bar{n}_{GSC}=\Gamma/\kappa$, where $\kappa=2/(30$~$\mu$s$)$ is the cooling rate. This heating is applied only during the sympathetic cooling/sideband step. \label{table:GSC}}
\end{centering}
\end{table*}

\subsubsection{Sensitivity to calibration errors}
\begin{figure*}
    \begin{centering}
	\includegraphics[]{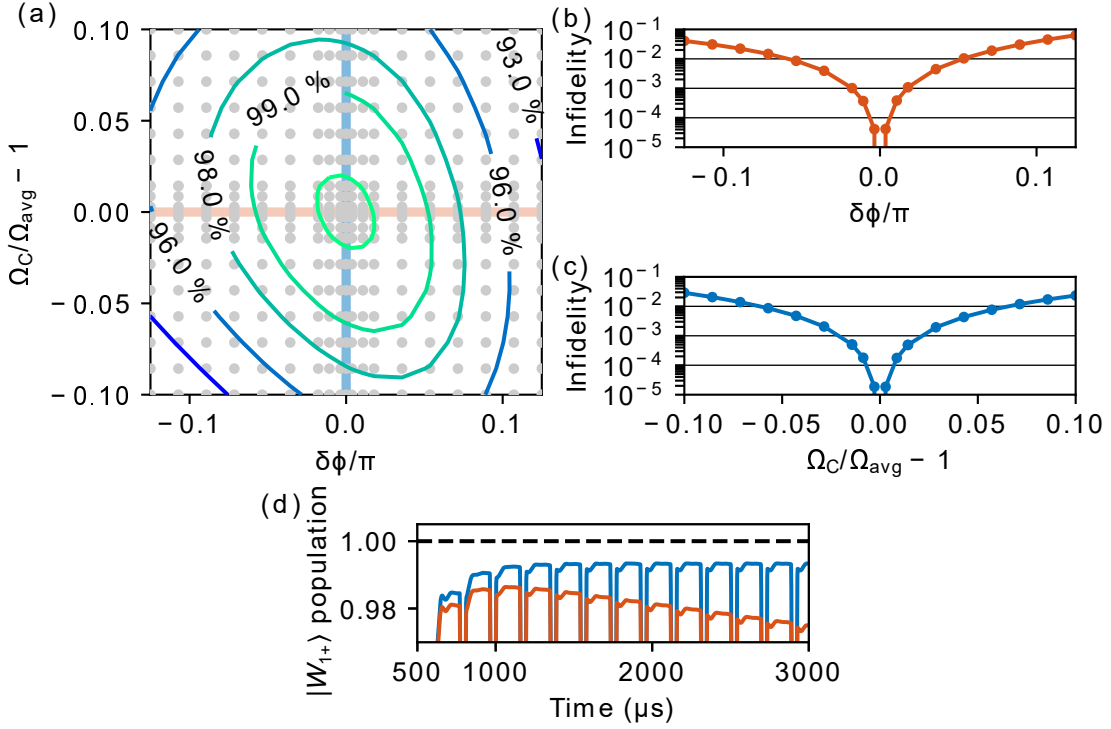}
	\caption{Investigation of sensitivity to calibration errors and photon scattering. (a)~Contour plot depicting target-state fidelity after twenty iterations as a function of phase error $\delta\phi$ and Rabi-rate balance error $\Omega_C/\Omega_{avg}-1$. The unlabeled center contour corresponds to 99.9 \% fidelity. Light gray points indicate points that were simulated for contour interpolation. Light blue and orange lines indicate cross-sections that are plotted in parts (b) and (c). (b, c)~Cross-sections of the contour surface plotted in part (a), showing infidelity as a function of phase error for no Rabi-rate mismatch (b), and as a function of Rabi-rate mismatch for no phase error (c). (d)~Simulated $\W{1+}$ population as a function of time with photon scattering error included as described in the text. Without repumping from leakage states (orange curve), the fidelity peaks at 98.6 \% after 1065 $\mu$s and then decays. With a repump time of 10 $\mu$s (blue curve, see text), the quasi-steady-state fidelity is 99.3 \%.} \label{fig:calibration_errors}
	\end{centering}
\end{figure*}

We investigate the protocol's sensitivity to calibration errors and present the results in Fig. \ref{fig:calibration_errors}. We consider failure to properly set the ion spacing, which results in interactions that do not neatly change the chirality according to $\delta s=0$ or $\pm1$, and failure to achieve uniformity of the sideband Rabi frequencies $|\Omega_j|$ across the ions. For simplicity we focus on the case of symmetrical errors. This covers several important failure mechanisms, for example imperfect adjustment of the depth of the potential well or of the strength of an even-order anharmonic term. We parameterize non-uniformity of the Rabi frequencies by the ratio $\Omega_C/\Omega_{avg}$ between the Rabi rate on the center qubit ion and the Rabi rate averaged over the qubit ions, with the rates for the left and right ions being the same. We parameterize ion-spacing errors by the deviation $\delta\phi$ of the phase difference between the ions of the Raman interaction. For example, for the $\delta s=-1$ sideband, the phases across the ions on the $a^\dag \sigma_-$ term in the Hamiltonian, which is relevant for generation of $\W{1+}$, are $-2\pi/3-\delta\phi$, $0$, and $2\pi/3+\delta\phi$. Since the MS gate can be made insensitive to this phase error by using the phase-insensitive geometry \cite{Lee2005}, and because in the phase-sensitive geometry the MS gate requires a different ion spacing than for the sideband-transition step, we simulate the phase error only on the sideband-transition step. As can be seen in Fig. \ref{fig:calibration_errors}, the fidelity is sensitive to these errors, but not prohibitively so.

\subsubsection{Choice of qubit states} Because this scheme does not use closed population cycling through an electronic excited state, there is more freedom in the choice for the pair of states that constitutes the qubit than for schemes that rely on spontaneous emission. A natural choice is a pair of states connected by a first-order magnetic-field-insensitive transition. This choice of a ``clock" transition has benefits such as reduced sensitivity to fluctuating magnetic field gradients, which otherwise reduce the scheme's fidelity.

As a concrete example, we envision using a qubit consisting of $\ket{F=2,m_F = 0}=\down$ and $\ket{F=1, m_F = 1} = \up$ within the \Be \Slevel manifold, which is first-order field-insensitive at an applied magnetic field of 11.96 mT \cite{Langer2005}.

\subsubsection{Infidelity due to spontaneous photon scattering} Spontaneous photon emission after excitation of a qubit ion to an electronic excited state is an important source of error in trapped-ion systems that use stimulated Raman transitions \cite{Ozeri2007}. To numerically investigate this effect, we simulate the case of \Be qubit ions addressed by laser beams red-detuned by 1 THz from the \Slevel$\leftrightarrow$\textsuperscript{2}P\textsubscript{1/2} transition, with beam power such that first-sideband $\pi$ times are 30 $\mu$s, as above. We assume an experimental geometry in which one Raman beam is aligned with the quantizing magnetic field at 45\textsuperscript{o} to the trap axis, and the other lies perpendicular to the first beam in the plane defined by the magnetic field and the trap axis, so that the wavevector difference between the two beams is axial. We choose $\hat{\sigma}_-$ polarization and lower frequency $\omega_-$ for the beam parallel to the magnetic field and $\hat{\pi}$ polarization and higher frequency $\omega_\pi>\omega_-$ for the beam perpendicular to it, so that $\omega_\pi-\omega_-=(E_\uparrow-E_\downarrow)/\hbar$ and resonant stimulated Raman transitions are driven between the $\up$ and $\down$ qubit states with energies $E_\uparrow$ and $E_\downarrow$, respectively. These polarization and frequency choices maximize the Raman Rabi rate and minimize the total scattering rate out of the $\up$ and $\down$ states at a given detuning from the $S_{1/2}\leftrightarrow P_{1/2}$ transition.

The model includes Raman scattering that leads to population loss from the qubit space by populating ground-state levels other than $\up$ and $\down$, as well as Raman scattering that leads to transitions between $\up$ and $\down$. Independently, these two errors lead to decay of the fidelity over time and to a reduction in the steady-state fidelity, respectively, since the latter can be corrected by the dissipative dynamics. The model also includes Rayleigh scattering, which induces decoherence between $\up$ and $\down$ and reduces the steady-state fidelity to the extent that the scattering amplitudes differ for the two qubit states.  Finally, we include the recoil associated with these scattering events, whereby they induce heating of the mode used in the scheme. We use a Lamb-Dicke parameter of $\eta = 0.24$ for these calculations, corresponding to the highest-frequency mode in the anharmonic potential shown in Fig. \ref{fig:MBBBMcrystal}, which has uniform participation of the qubit ions. The calculations are conducted for an applied quantization field of 11.96 mT, as discussed above.

The results of the simulation are shown in Fig. \ref{fig:calibration_errors}d. In this simulation a maximum fidelity of 98.6 \% is reached after 1065 $\mu$s of operation, and then the fidelity decays due to loss of population from the qubit space. Since this scheme uses iterated dissipative dynamics to populate the target state, population leakage can be corrected by using repumper lasers that drive population back to e.g. $\down$. In the limit of instantaneous repumping, population leakage error is then converted into decay from $\up$ to $\down$ and dephasing of the $\down$ state. For a reasonable repumping time of $\tau_{rep} = 10$ $\mu$s from each leakage state (where the corresponding collapse operator is scaled by $\sqrt{1/\tau_{rep}}$), we find a quasi-steady-state fidelity of 99.3 \%. The model includes heating due to recoil from scattering on repump transitions, but it neglects off-resonant driving of other transitions by the repump beams.

Error due to spontaneous photon scattering is qubit-species dependent \cite{Ozeri2007}. In general, independent of this choice, the rate of spontaneous Raman scattering can be decreased while maintaining the rate of the desired stimulated transitions by increasing the laser power and detuning. The Rayleigh scattering rate cannot be reduced by the same approach, but the decoherence effect of Rayleigh scattering is typically quite small. For example, under the conditions discussed above we calculate the rate of the decoherence-inducing differential Rayleigh scattering to be $\sim2\times10^{-4}$ times the total leakage rate. On the other hand, the total Rayleigh scattering rate is $\sim174$ times the total leakage rate, but this effect leads only to heating through recoil, not to state changes or decoherence. Finally, it would be possible to entirely eliminate photon scattering error if all sideband transitions were driven by a different mechanism, such as a combination of microwave radiation and magnetic-field gradients \cite{Mintert2001, Johanning2009, Warring2013, Srinivas2019}, but this would come with its own technical challenges.

\section{Discussion\label{sec:discussion}}

Using the couplings available in trapped-ion systems, we have designed dissipative protocols for the preparation of W states of three ions. The most promising protocol combines unitary couplings in the form of sideband interactions and MS gates with dissipation through sympathetic cooling, and it avoids the need for addressing of individual ions. The resulting effective dissipation is engineered through cooling of motional excitations, i.e. of an external degree of freedom. Despite being dissipative, this interaction allows correlations between internal degrees of freedom, the qubits, to survive. Damping of harmonic oscillator auxiliary subsystems \cite{Kastoryano2011, Lin2013, Reiter2017} or coupled systems \cite{Barreiro2011} has previously allowed for the implementation of dissipative schemes, and this approach holds promise for future protocols extending the range of dissipative quantum computation. Moreover, the application of the MS gate for dissipative preparation of W states shows again that time-dependent interactions can be useful to aid the generation of quasi-steady states, as previous work has demonstrated \cite{Barreiro2011, Lin2013}.

The protocols we have presented are part of a new generation of strategies for dissipative entanglement preparation that eschew timescale hierarchies in favor of symmetry-based dissipation engineering. As a consequence, the fidelities predicted by numerical simulations are comparable to those achievable with unitary schemes \cite{Lin2016a}. Future efforts may explore these concepts beyond state preparation, such as in new protocols for autonomous quantum error correction and quantum sensing \cite{Reiter2017}, as well as for quantum simulation \cite{Raghunandan2020, Reiter2019}.

\begin{acknowledgements}
The authors thank Alejandra Collopy and Nathan Lysne for helpful comments on the manuscript and Emanuel Knill for helpful discussions. D. C. C. acknowledges support from a National Research Council postdoctoral fellowship. S. D. E. acknowledges support from the National Science Foundation under grant DGE 1650115. F. R. acknowledges financial support from the Swiss National Science Foundation (Ambizione grant no. PZ00P2$\_$186040).  P.-Y. H and J. J. W. acknowledge support from the Professional Research Experience Program (PREP) operated jointly by NIST and University of Colorado Boulder.
\end{acknowledgements}

\appendix

\section{Achieving Uniform Rabi Rates Using Anharmonicity\label{app:crystal_modes}}
As discussed in the main text, the fidelity achieved by the protocols we present is maximized when the Raman beam intensity is equal on the qubit ions and the qubit ions have uniform participation in the driven mode. Ideally, this mode also has only weak coupling to electric field noise that is homogeneous over the length of the ion chain (in order to reduce heating) and is separated in frequency from other modes to avoid off-resonant coupling. For a linear ion crystal these requirements can be satisfied by adjusting the axial trapping potential, which affects the ion equilibrium positions and mode vectors. Specifically, we consider electrostatic trapping voltages of the form:
\begin{equation}
V(x) = \frac{1}{2}k_2 x^2 + \frac{1}{4}k_4 x^4,\label{eq:anharmonicpotential}
\end{equation} 
which represents the simplest symmetrical deviation from a harmonic potential. By adjusting the terms $k_2$ and $k_4$, modes with the desired relative ion participation can be engineered.

With three qubit ions, the smallest symmetrical mixed-species crystals have two auxiliary ions and always have a qubit ion in the center. Modes with even parity (e.g. $(\leftarrow,\leftarrow,0,\rightarrow,\rightarrow)$, schematically) do not couple to uniform electric fields and consequently have very low heating rates, but they also have no participation from the center ion. In a crystal with ions of non-uniform masses, the odd-parity modes typically have some center-of-mass motion and therefore couple to homogeneous electric fields. The heating rate $\dot{\bar{n}}_l$ of a given mode $l$ with frequency $\omega_l$ due to excitation by a uniform external field at the same frequency is given by (generalized from Ref. \cite{Kielpinski2000}):
\begin{align}\label{eq:heatingrate}
\dot{\bar{n}}_l&=\frac{q^2 S_E(\omega_l)}{4\hbar\omega_l}\left(\sum_{j=1}^N\frac{ v_j^{(l)}}{\sqrt{m_j}}\right)^2,\\
&=\frac{q^2 S_E(\omega_l)}{4\hbar\omega_l} C^{(l)},
\end{align}
where $q$ is the fundamental charge, $S_E(\omega_l)$ is the power spectral density of spatially uniform electric-field noise, $v^{(l)}$ is the (root-mass-weighted) normalized eigenvector for mode $l$, and $m_j$ are the ion masses. We define $C^{(l)}$ as a frequency-independent measure of the mode's coupling to global fields. Typically, a given mode's heating rate may be decreased by increasing the mode's frequency.

By considering the axial trapping potential and the Coulomb force, the equilibrium positions of the ions and the normal mode vectors $v^{(l)}$ can be numerically determined \cite{Brown2011a, James1998}. In order to investigate the general practicality of the schemes we described in the main text, we performed this analysis for several mixed-species ion crystal configurations.
	
We considered four mixed-species ion combinations: \Be/\Mg, \Be/\textsuperscript{40}Ca\textsuperscript{+}\xspace,  \textsuperscript{40}Ca\textsuperscript{+}\xspace/\textsuperscript{88}Sr\textsuperscript{+}\xspace, and \textsuperscript{138}Ba\textsuperscript{+}\xspace/\textsuperscript{171}Yb\textsuperscript{+}\xspace. We also consider an equal-mass, mixed-state configuration of \textsuperscript{43}Ca\textsuperscript{+}\xspace where ions of the same species but in a different state act as coolant ions \cite{Moore2020}. We explored this range of masses and mass ratios to illustrate the effect mass has on the feasibility of achieving ideal engineered modes for this experiment. Engineering modes with equal ion participation where the frequencies of all five axial modes are well-separated ($>$100s of kHz away) is particularly difficult for the \textsuperscript{138}Ba\textsuperscript{+}\xspace/\textsuperscript{171}Yb\textsuperscript{+}\xspace combination due to the inherently low mode frequencies of these heavier ions. For all the ions we consider, the value of $k_2$ varies from $30$ to $60$ $\mu$V/$\mu$m$^2$ for typical single-ion harmonic trapping potentials \cite{Wright2015, Bruzewicz2019, Negnevitsky2018}.

The values of $k_4$ that can be practically realized depend on geometric factors such as the size of the electrodes and distance from the ion, as well as the maximum electrode voltage that the system can achieve \cite{Home2006}. We used a maximum anharmonicity of $k_4\approx0.0156$ $\mu$V/$\mu$m$^4$ because this should be achievable in traps such as those described in Refs. \cite{Blakestad2010, Brown2011a}. In general, ion traps with smaller ion-to-electrode distance and higher voltages should be capable of higher $k_4$.

Most of the ion chains we considered are symmetric combinations in the order $m_L m_H m_H m_H m_L$, where $m_L$ ($m_H$) denotes the lighter (heavier) ion of the two. We also investigated crystals of the form $m_L m_H m_L m_H m_L$, and we found that in general the modes were not suitable when tuned to have equal ion participation in each species: using the highest frequency mode, the difference between the highest two frequencies was consistently less than 4\%; the middle frequency mode has low $m_H$ ion participation; and the lowest frequency mode has high heating rate. With the exception of the MBBBM crystal presented in the main text (which is preferred over BMMMB because \Be has a lower spontaneous emission error than \Mg for a given laser intensity, so it serves as a better qubit), we do not consider crystals of the form $m_H m_L m_L m_L m_H$ when a good set of parameters can be found for other configurations, because these may be more difficult to arrange in practice.

Table \ref{table:mode_results} summarizes the results of the analysis. We present the normal mode participation vectors and frequencies for optimized values of $k_2$ and $k_4$, along with the mode's coupling to homogeneous external fields as defined in Eq. \ref{eq:heatingrate}. For the participation vector, we list $z^{(l)}_j = \sqrt{\hbar/2m_j\omega_l} v^{(l)}_j$, the ground state wavefunction root-mean square size for ion $j$ in mode $l$ \cite{Home2011}. The Lamb-Dicke parameter $\eta_j^{(l)}=\Delta k \cdot |z^{(l)}_j|$ quantifies ion $j$'s coupling in mode $l$ to a pair of Raman beams with axial wavenumber difference $\Delta k$. The heating rate parameter $C^{(l)}$ is heavily dominated by the ion mass, so we normalize it by the $C_0$, the value of $C^{(l)}$ for the in-phase mode of the same crystal in a harmonic potential (which is the closest to a ``true" center-of-mass mode for a mixed-species ion crystal in a harmonic potential, and thus has the highest heating rate).

\begin{table*}
\begin{centering}
	\begin{tabular}{|c|c|c|c|c|}
		\hline
		Ion Chain & \shortstack{Normal-Mode Amplitudes\\$z^{(l)}_j$ (nm)} & \shortstack{Freq.\\(MHz)} & \shortstack{$k_2$\\($\mu$V/$\mu$m$^2$)} &$C^{(l)}/C_0$\\\hline
		Mg-\textbf{Be}-\textbf{Be}-\textbf{Be}-Mg* & -1.38, \textbf{8.35}, \textbf{-8.35}, \textbf{8.35}, -1.38 & 2.56 & -0.729 & 0.0542\\\hline
		Be-\textbf{Mg}-\textbf{Mg}-\textbf{Mg}-Be & -4.17, \textbf{2.66}, \textbf{-2.66}, \textbf{2.66}, -4.17 & 6.00 & 25.0 $^{(-)}$ & 0.153\\\hline
		Ca-\textbf{Be}-\textbf{Be}-\textbf{Be}-Ca* & -0.823, \textbf{8.58}, \textbf{-8.58}, \textbf{8.58}, -0.823 & 2.48 & -0.894 & 0.123\\\hline
		Ca-\textbf{Ca}-\textbf{Ca}-\textbf{Ca}-Ca** & -2.74, \textbf{4.74}, \textbf{-4.74}, \textbf{4.74}, -2.74 & 1.42 & 0.436 & 0.001\\\hline
		Ca-\textbf{Sr}-\textbf{Sr}-\textbf{Sr}-Ca & -3.53, \textbf{2.70}, \textbf{-2.70}, \textbf{2.70}, -3.53 & 1.73 & 6.12 & 0.093\\\hline
		Ba-\textbf{Yb}-\textbf{Yb}-\textbf{Yb}-Ba*** & -2.32, \textbf{3.13}, \textbf{-3.13}, \textbf{3.13}, -2.32 & 0.778 & 1.05 & 0.011\\\hline
		Ba-\textbf{Yb}-\textbf{Yb}-\textbf{Yb}-Ba & 3.45, \textbf{3.51}, \textbf{3.51}, \textbf{3.51}, 3.45 & 0.527 & 19.2 $^{(-)}$ & 1.00\\\hline
	\end{tabular}
	\caption{Sample ion chain configurations for various species combinations along with $k_2$ parameter that achieves uniform coupling amplitudes for the qubit ion (bold). All $k_4$ parameters used are of amplitude 0.0156 $\mu V/\mu m^4$ (an ``experimental maximum" that we enforce, although it is very likely that many systems can achieve higher amplitudes), with the exponent $^{(-)}$ on the corresponding $k_2$ parameter when the value of $k_4$ used was negative. $C^{(l)}/C_0$ quantifies the coupling to spatially uniform electric field noise compared to the coupling for the lowest-frequency mode of the same crystal in a harmonic potential. Unless otherwise stated, crystals in configurations listed above have a $>$130 kHz separation between nearest frequency modes, ensuring off-resonant coupling is unlikely. *Crystals with lighter ions in the center may be more difficult to arrange, depending on the system's capability to rearrange ions. **Here we use ``Ca" to mean  \textsuperscript{43}Ca\textsuperscript{+}\xspace. All other ``Ca" in this table refer to \textsuperscript{40}Ca\textsuperscript{+}\xspace. ***In this configuration, the next highest frequency mode is at 0.737 MHz, so the scheme may suffer from significant off-resonant coupling. If the experimental setup is capable of higher $k_4$ and $k_2$ values, the frequencies, as well as the frequency difference, can be increased. For this ion chain we list the in-phase mode in the next row, which does not suffer from this issue but has a higher heating rate.\label{table:mode_results}}
	\end{centering}
\end{table*}

\bibliography{DissipativeWStates}

\end{document}